\documentclass[twocolumn,epsfig,APS,floats]{revtex4}
\usepackage{graphicx}
\usepackage{rotating}
\usepackage{color}

\usepackage{fancyhdr}
\fancypagestyle{titlepage}{
\fancyhead[R]{J. Phys. Chem. C. \textbf{120}, 12890 (2016)}}
\begin{document}


\title{Structure and Bonding in Amorphous Cr$_{1-x}$C$_{x}$ Nanocomposite Thin Films: X-ray Absorption Spectra and First-Principles Calculations}

\author{Weine Olovsson$^1$, Bj\"{o}rn Alling$^{2,3}$, and Martin Magnuson$^2$}

\affiliation{$^1$Theoretical Physics, Department of Physics, Chemistry and Biology (IFM), Link\"{o}ping University,
SE-581 83 Link\"{o}ping, Sweden}

\affiliation{$^2$Thin Film Physics Division, Department of Physics, Chemistry and Biology (IFM), Link\"{o}ping University,
SE-581 83 Link\"{o}ping, Sweden}

\affiliation{$^3$Max-Planck-Institut f\"{u}r Eisenforschung GmbH, D-402 37 D\"{u}sseldorf, Germany}

\begin{abstract}
The local structure and chemical bonding in two-phase amorphous Cr$_{1-x}$C$_{x}$ nanocomposite thin films are investigated by Cr $K$-edge ($1s$) X-ray absorption near-edge structure (XANES) and extended X-ray absorption fine structure (EXAFS) spectroscopies in comparison to theory. By utilizing the computationally efficient \textit{stochastic quenching} (SQ) technique, we reveal the complexity of different Cr-sites in the transition metal carbides, highlighting the need for large scale averaging to obtain theoretical XANES and EXAFS spectra for comparison with measurements. As shown in this work, it is advantageous to use \textit{ab initio} theory as an assessment to correctly model and fit experimental spectra and investigate the trends of bond lengths and coordination numbers in complex amorphous materials. With sufficient total carbon content ($\geq$ 30 at\%), we find that the short-range coordination in the amorphous carbide phase exhibit similarities to that of a Cr$_{7}$C$_{3\pm{}y}$ structure, while excessive carbons assemble in the amorphous carbon phase.
\end{abstract}

\maketitle

\section{Introduction}
In the quest for new coating materials with outperforming corrosion and wear resistance properties, chromium carbide thin films are of high interest in applications for electrical contacts and decorative purposes~\cite{ Cavaleiro,Cheng}. Metal carbide films produced by physical vapor deposition (PVD), show different microstructures, ranging from epitaxial single-crystal materials to nanocomposites with crystallites of the carbide in an amorphous carbon/amorphous carbide matrix~\cite{Vieira1,Vieira2}.
Completely amorphous metal carbide coatings are also found, in both binary and ternary systems such as Cr-C, Fe-C, Ni-C, and W-Fe-C~\cite{Cavaleiro,Furlan2014,Furlan2015}. It is an important methodological challenge to theoretically model such complex non-crystalline environments, in order to understand their structure and bonding properties.

In the case of {\it crystalline} materials, there is a multitude of successful combined experimental and theoretical studies in the literature, see for example Refs.~\cite{Magnuson2004,Magnuson2008,Magnuson2008b,Mizoguchi2010,Olovsson2011,Olovsson2013}. On the contrary, it is far from straightforward to model {\it amorphous} structures for studying materials properties. The lack of long-range translational symmetry provides a formidable challenge for obtaining suitable model structures of the systems. This is also the case experimentally, as the samples become {\it X-ray amorphous} with no usable diffraction peaks to determine the structure. Furthermore, in the computation of XANES and EXAFS spectra, the absence of symmetry means that the excitation response of each atom is unique and can be very distinct. 
As spectral measurements are effective samplings of individual excitations of atoms in different local chemical environments, an average over a large number of sites must be considered in a theoretical approach. While detailed explanation for the formation of non-crystalline phases during PVD in certain carbide systems remains unclear, it involves a significant quenching for conditions with neutral atoms and substrate temperatures below the melting point of the carbide, giving rise to limited surface mobility where atoms only find locally favorable sites. It is known that amorphous films often occur in metallic carbide systems where crystalline phases exhibit structural units that have carbon atoms in a mixture of both octahedral and prismatic sites~\cite{Bauer-Grosse1}. 

Previous experimental studies of amorphous Cr$_{1-x}$C$_{x}$ films~\cite{Gaskell,Bauer-Grosse2,Music,Jiang,Andersson2012} suggest that the films contain two components; an amorphous metal-rich carbide phase, and an amorphous carbon-rich phase. With increasing total carbon content, the amorphous carbon-rich contribution is believed to grow and incorporate the additional carbon, leaving the metal-rich phase more or less constant in carbon content ~\cite{Gaskell,Bauer-Grosse2,Music,Jiang,Andersson2012,Magnuson1}. 
From C $1s$ X-ray photoelectron spectroscopy (XPS) measurements on the same samples as in this work, it was determined, based on quantitative estimations of carbide and a-C contributions in the spectra, that the amorphous metal carbide phase contained approximately 20-30 at.\% of carbon. This is in the same compositional range as e.g.\ the crystalline Cr$_{23}$C$_{6}$ and Cr$_{7}$C$_{3}$ phases.
On the other hand, the amorphous carbon phase contained approximately pure carbon with a high amount of $sp^{2}$-hybridized graphene-like carbon for large amounts of total carbon contents~\cite{Magnuson2}. 
However, to identify the detailed structure and the bonding of the carbide phase in these complex non-equilibrium materials~\cite{Magnuson2}, more detailed experimental and theoretical studies are needed, in particular focusing on the surrounding of the metal atoms.

In this work, we investigate and compare experiments for nanocomposite Cr$_{1-x}$C$_{x}$ thin films with theory for {\it completely amorphous} systems as a function of carbon concentration, together with {\it ideal crystal} structures.
X-ray absorption near-edge structure (XANES) and extended X-ray absorption fine structure (EXAFS) spectroscopies are ideal tools for characterizing the structure and chemical bonding in crystalline materials~\cite{Cockayne,Jansson}.
Since these techniques are element specific, we anticipate that they can also be applied as efficient probes of the local environment in more complex non-crystalline materials. In particular, we investigate the Cr $K$ near-edge ($1s$) and EXAFS signals of the CrC$_x$ carbide phase in amorphous Cr$_{1-x}$C$_{x}$ thin films by experiment and {\it ab initio} calculations, including metal Cr for comparison. In order to model {\it completely} amorphous systems, we apply the so-called {\it stochastic quenching} (SQ) method~\cite{Kadas,Holmstrom1,Holmstrom2}. 

\section{Theory}

\subsection{Structural models}
All structural models were obtained using density functional theory (DFT)~\cite{Hohenberg1964,Kohn1965} utilizing the projector augmented wave (PAW)~\cite{Blochl1994} method as implemented in the Vienna ab-initio simulation package ({\tt VASP})~\cite{Kresse1,Kresse2}. The generalized gradient approximation (GGA)~\cite{PBE1996} for electronic exchange and correlation effects was used in these simulations. The atoms are first placed randomly in a supercell and then relaxed using DFT until the force on each atom is negligible~\cite{Kadas,Holmstrom1,Holmstrom2}. 
In this way a large set of structures is obtained, which together represent a specific concentration, at a comparatively low computational cost.
Finally, to produce the theoretical X-ray absorption spectra, an average is taken over spectra computed at each Cr-atom site in the amorphous structure.
In order to consider possible microstructures and phases within an amorphous matrix, we also consider crystalline CrC$_{x}$ structures that exist experimentally: \emph{cF}116-Cr$_{23}$C$_{6}$, \emph{oP}16-Cr$_{3}$C, \emph{hP}20-Cr$_{7}$C$_{3}$, \emph{hP}80-Cr$_{7}$C$_{3}$, \emph{oP}40-Cr$_{7}$C$_{3}$, \emph{oP}20-Cr$_{3}$C$_{2}$, as well as the cubic B1 (octahedral) and hypothetical hexagonal hP2 (trigonal prismatic) type of CrC.
All spectra are calculated by employing the computationally efficient real-space Green's function code {\tt FEFF9}~\cite{Rehr2000,Rehr2009,Rehr2010}.
In this way it is possible to produce the many individual spectra which are needed to represent amorphous structures. Different magnetic structures, ferromagnetic and antiferromagnetic were considered as initial configurations, although, with the exception of B1 CrC, magnetic moments relaxed to zero or almost zero during the electronic structure calculation. 

The structural models used for simulations of the spectra of amorphous Cr$_{0.67}$C$_{0.33}$, Cr$_{0.5}$C$_{0.5}$, and Cr$_{0.33}$C$_{0.67}$ were obtained using the stochastic quenching approach~\cite{Holmstrom1,Holmstrom2,Kadas} in combination with DFT.  Cubic supercells containing 250 atoms (Cr$_{0.5}$C$_{0.5}$) or 252 atoms (Cr$_{0.67}$C$_{0.33}$ and Cr$_{0.33}$C$_{0.67}$) were used, allowing for the selected concentrations to be modeled accurately. The stochastic quenching procedure was carried out in a similar procedure as in Ref.~\cite{Kadas}. The atomic positions were initiated with a random distribution in the simulation box, avoiding, for numerical stability concerns in the electronic structure calculations, pair distributions shorter than 0.8 \AA, which is about 0.6 times the calculated C-C dimer distance in vacuum. 
This random structure is then quenched by gradually relaxing the internal atomic positions inside the supercell, until they reach locally stable positions with forces on each ion being smaller or in the order of 0.02 eV/\AA.  

For each composition, five or six volumes, with 19-25 SQ structures per volume, were used to obtain the equilibrium volume with a cubic polynomial fit. A similar number of SQ structures were then used to obtain average pair distribution functions g(r) for each bond type and to simulate spectroscopic properties. The calculations were initiated with individual Cr atoms having either spin up or spin down magnetic polarization with equal probability and a spatially random distribution in order to simulate a paramagnetic configuration. Some Cr atoms, in particular in the compositions Cr$_{0.33}$C$_{0.67}$, retained a magnetic moment after the SQ procedure with a mean magnitude of about 0.8 $\mu_{B}$ with minimum and maximum magnitudes 0.1 and 2.2 $\mu_{B}$, respectively. For the composition Cr$_{0.5}$C$_{0.5}$ the corresponding mean value is about 0.3 $\mu_{B}$ with a few of the Cr moments being around 2 $\mu_{B}$. For the Cr$_{0.67}$C$_{0.33}$ composition most of the local moments disappear with a mean magnitude of around 0.02 $\mu_{B}$. 
Overall, the local magnetic moments are clearly less robust as compared to similar simulations for the related amorphous CrN-system~\cite{Lindmaa2013}.
A Gaussian broadening of the calculated pair distances of 0.2~\AA\ was used in the simulations.
Figure 1 shows one of the supercells (SQ1) generated by the procedure of the SQ method in order to model amorphous Cr$_{0.5}$C$_{0.5}$. 
Large blue spheres correspond to Cr and smaller brown spheres are C atoms. In order to obtain the theoretical X-ray absorption spectrum for an amorphous system, 
an average of all the spectra for the different SQ supercells and Cr sites was performed.
\begin{figure}
\includegraphics[width=90mm]{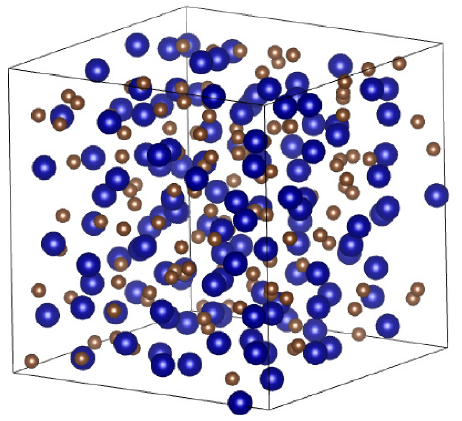} 
\vspace{0.2cm} 
\caption[] {One of the structures (SQ1) obtained by the stochastic quenching method for amorphous Cr$_{0.5}$C$_{0.5}$. The Cr-atoms are represented
by large blue spheres, while C-atoms are shown as smaller brown spheres. The supercell was drawn using {\tt VESTA}~\cite{vesta}.}
\label{fig1}
\end{figure}

\subsection{X-ray absorption spectra}
In order to produce theoretical XANES  and EXAFS spectra for the structures described above, we employ the single particle Green's function technique within
DFT~\cite{Hohenberg1964,Kohn1965}.
All calculations were performed utilizing the computationally efficient {\it ab initio} real-space Green's function (RSGF) approach of the {\tt FEFF9} code~\cite{Rehr2000,Rehr2009,Rehr2010}.
For example, in the case of Cr$_{0.5}$C$_{0.5}$ the resulting Cr $K$-edge spectrum is an average over 22 SQ structures, each in turn an average over 125 Cr sites,
amounting to 2750 spectra in total to represent the amorphous system.
All the potentials are calculated self-consistently and the default Hedin-Lundquist exchange correlation potential is used.
In more detail, to obtain accurate near-edge structures considering XANES, we utilized the real-space full multiple scattering (RSFMS) approach within {\tt FEFF9}.
Here, the potentials were calculated for the central scattering Cr atom and for the other Cr and C atoms.
To take the final state effect into account, a fully screened core-hole was used for the central scattering atom, i.e.\ applying the final-state rule, or core-hole approximation.
Clusters with more than 550 atoms were employed in order to converge the spectral features in the different systems.
For a brief overview of different methodologies for calculating XANES see Ref.~\cite{Mizoguchi2010} and references within.
For EXAFS, which considers the energy region far above the near-edge, it is sufficient to use multiple scattering paths with no approximation for the core-hole~\cite{Rehr2000}.
Finally, the theoretical spectra were aligned with experiment by applying a rigid shift.

\section{Experimental}
Amorphous chromium carbide (Cr$_{1-x}$C$_{x}$) samples with different carbon content were made using direct current (DC) magnetron sputtering as described elsewhere~\cite{Andersson2012,Magnuson2}. 
It was also found~\cite{Andersson2012} that the samples contain about 2-5 at.\% oxygen present.
The XANES and EXAFS spectra were measured at the undulator beamline I811 on the MAX II ring of the MAX IV Laboratory, Lund University, Sweden~\cite{811}. The energy resolution at the Cr $1s$ edge of the beamline monochromator was better than 0.5 eV. The X-ray absorption spectra were recorded in reflection mode by collecting the fluorescence yield using a PIPS detector from -150 eV below to 1000 eV above the Cr $1s$ absorption edge (5989 eV) with 0.5 eV energy steps. 
To avoid self-absorption effects in the sample and Bragg peaks from the substrate, the incidence angle on the sample was varied in 0.25 degrees steps in a maximum range of $\pm$3 degrees from normal incidence with a step motor. 

Based on the XANES fitting results (described in the next Section), two Cr-C and two Cr-Cr scattering paths obtained from {\tt FEFF9}~\cite{Rehr2000,Rehr2009,Rehr2010} were included in the EXAFS fitting procedure with \emph{hP}20-Cr$_7$C$_3$ as a model system using the {\tt VIPER} software package~\cite{Klementev2001}. The reference energy $E_{0}$ was set to the pre-peak of each Cr $1s$ X-ray absorption spectrum as determined from the first peak of the derivative of each spectrum relative to the pure bcc-Cr reference sample. In the fitting procedure, $E_0$ was set to 5988.6 eV (25\% C), +0.1 eV (42 and 47\%, as compared to the 25\% C sample), +0.4 eV (67\% C) and +0.6 eV (85\% C), respectively. 

The EXAFS functions $\chi k^2$ were extracted from the raw absorption data after removing remaining glitches originating from substrate diffraction peaks, subsequent background subtraction, and summarizing and averaging of 15 absorption spectra. The bond distances (R), number of neighbors (N), Debye-Waller factors ($\sigma^2$, representing the amount of disorder) and $\chi^{2}_{r}$=reduced $\chi^2$ as the squared area of the residual, were determined by fitting the back-Fourier-transform signal between $k$=0-12 \AA{}$^{-1}$ obtained from the Fourier-transform within $R$=0.7-3 \AA{} of the first coordination shell using a Hanning window function )~\cite{Egami} and a global electron reduction factor of $S_0^2$=0.8. The calculated pairs of Cr-C and Cr-Cr 1st and 2nd scattering paths in the model were obtained using the {\tt FEFF9} code~\cite{Rehr2000,Rehr2009,Rehr2010}.

\section{Results and discussion}
In this Section, we start by considering theoretical XANES and EXAFS Cr $K$-edge spectra for the amorphous and crystal chromium carbide structures, with a comparison to measurements carried out for the thin film systems and pure Cr metal. An analysis of the XANES data was done by fitting the measured spectra with the calculations using a  superposition of an amorphous phase and a smaller contribution of different crystalline phases.
Using the results of the XANES fitting, an analysis of the EXAFS data was carried out for the experimental data using a standard fitting procedure for crystalline materials, in order to obtain reduced (radial) pair distribution functions (RDFs), 
that are compared to theoretical RDF results.
Finally, the trend in the structural parameters of bond lengths and coordination numbers extracted from the EXAFS analysis are compared and discussed with theoretical values.

\begin{figure}
\includegraphics[width=90mm]{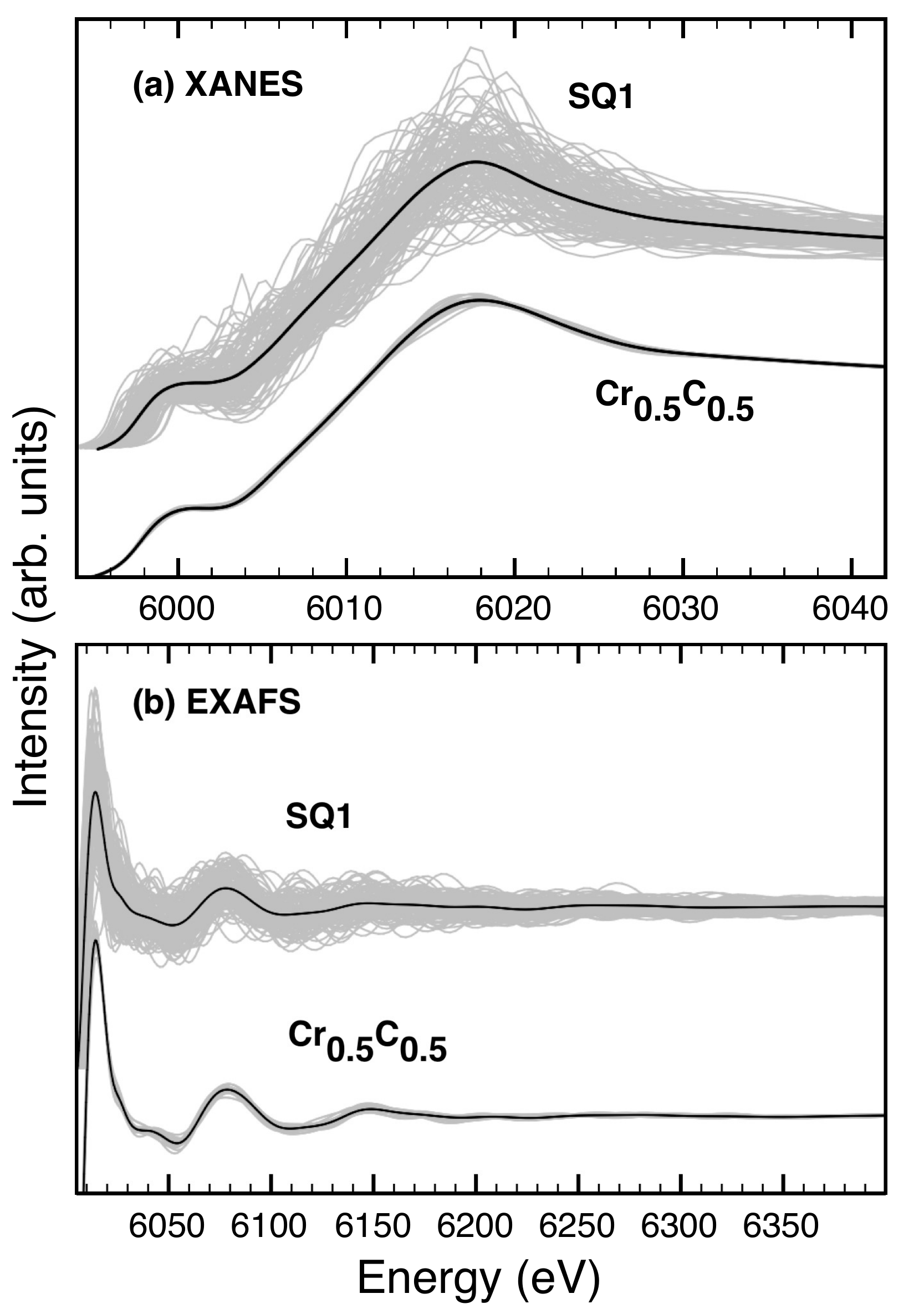} 
\vspace{0.2cm} 
\caption[] {Theoretical X-ray absorption spectra at the Cr $1s$-edge, a) XANES and b) EXAFS, 125 Cr-sites (gray lines), are exemplified for the structure SQ1 in Figure 1, together with the
spectra for Cr$_{0.5}$C$_{0.5}$ over all 22 SQ structures. The average of the spectra are shown as black lines.}
\label{fig2}
\end{figure}

In Figure 2, the large differences of the theoretical a) XANES and b) EXAFS spectra in the case of Cr$_{0.5}$C$_{0.5}$ are shown. The upper parts of the two panels a) and b) give the Cr $K$-edge spectra calculated over the 125 different Cr-sites (gray lines) for the SQ1 structure illustrated in Fig.\ 1. The averaged spectra are shown as solid black lines. The lower parts of the panels correspond to the averaged results obtained for each of the 22 SQ structures (gray lines). Finally, the resulting XANES and EXAFS spectra over the whole amorphous Cr$_{0.5}$C$_{0.5}$ structure are given by the solid black lines in the lower part of each panel. As observed in Fig.\ 2, the X-ray absorption fine structures are highly sensitive to the atom local environment in the amorphous matrix. Similar behavior was found at the other concentrations.

Our result in Fig.\ 2 contains an important message to theoreticians working with spectroscopical simulations of disordered materials, and in particular amorphous substances. It is clear that in order to give even a qualitatively reasonable description of the Cr XANES or EXAFS spectra, averaging over hundreds of atomic sites is needed. However, for a quantitative simulation of the average signal, which is what is measured experimentally, in particular for the EXAFS case it might not be enough to consider just one amorphous supercell. It can be seen in the lower curve of Fig.\ 2 that the computed average signals for some of the SQ supercells deviates considerably from the average line of all the samples. Thus, it might not be reliable to simulate EXAFS spectra with just one supercell, say generated with molecular dynamics and containing one- or two hundred atoms, even if the average spectra of all atoms in the supercell is considered. 

\begin{figure}
\includegraphics[width=90mm]{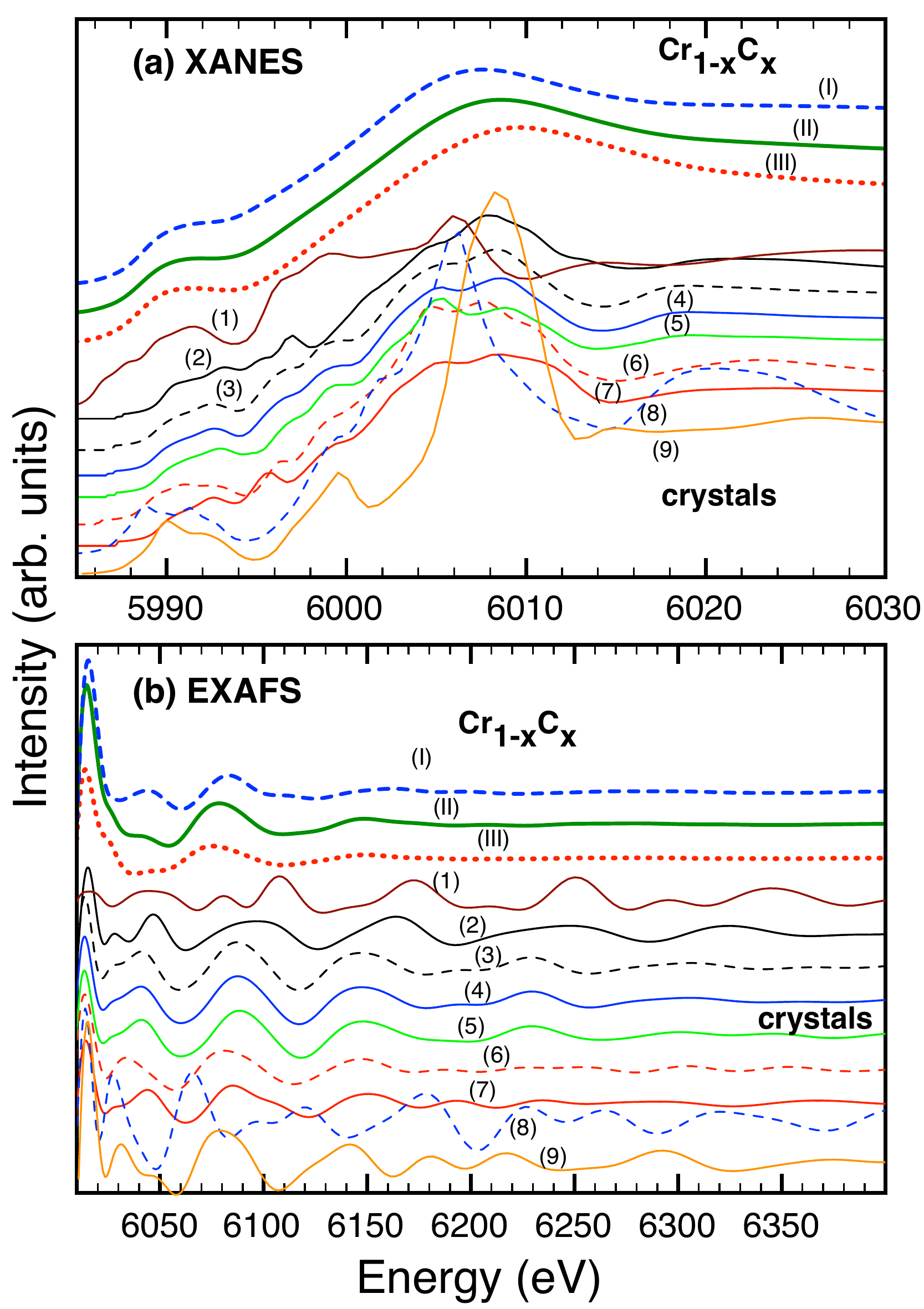} 
\vspace{0.2cm} 
\caption[] {(Theoretical X-ray absorption spectra for the Cr $1s$-edge, a) XANES and b) EXAFS, are compared for amorphous Cr$_{0.67}$C$_{0.33}$ (I),
Cr$_{0.5}$C$_{0.5}$ (II), Cr$_{0.33}$C$_{0.67}$ (III) and with the different crystal structures:
bcc-Cr (1), \emph{cF}116-Cr$_{23}$C$_{6}$ (2), \emph{oP}40-Cr$_{7}$C$_{3}$ (3), \emph{hP}80-Cr$_{7}$C$_{3}$ (4), \emph{hP}20-Cr$_{7}$C$_{3}$ (5),
\emph{oP}20-Cr$_{3}$C$_{2}$ (6), \emph{oP}16-Cr$_{3}$C (7), phases and cubic B1 [NaCl] (8) and hexagonal hP2 [WC] (9) type CrC.}
\label{fig3}
\end{figure}

In Figure 3, a comparison of theoretical a) XANES and b) EXAFS spectra is made between amorphous Cr$_{1-x}$C$_{x}$ and crystalline CrC$_x$ structures. In the upper parts of each panel, results for three different amorphous Cr$_{1-x}$C$_{x}$ structures with concentrations $x=0.33$ (blue dashed lines), $0.5$ (green line) and $0.67$ (red dotted line) are shown. In the lower part of each panel, the corresponding spectra for the crystals are presented: bcc-Cr, \emph{cF}116-Cr$_{23}$C$_{6}$, \emph{oP}16-Cr$_{3}$C, \emph{hP}20-Cr$_{7}$C$_{3}$, \emph{hP}80-Cr$_{7}$C$_{3}$, \emph{oP}40-Cr$_{7}$C$_{3}$, \emph{oP}20-Cr$_{3}$C$_{2}$, structures from experiment and cubic B1 (octahedral NaCl) and hypothetical hexagonal  \emph{hP}2 (trigonal prismatic WC) type of CrC.

\begin{figure}
\includegraphics[width=90mm]{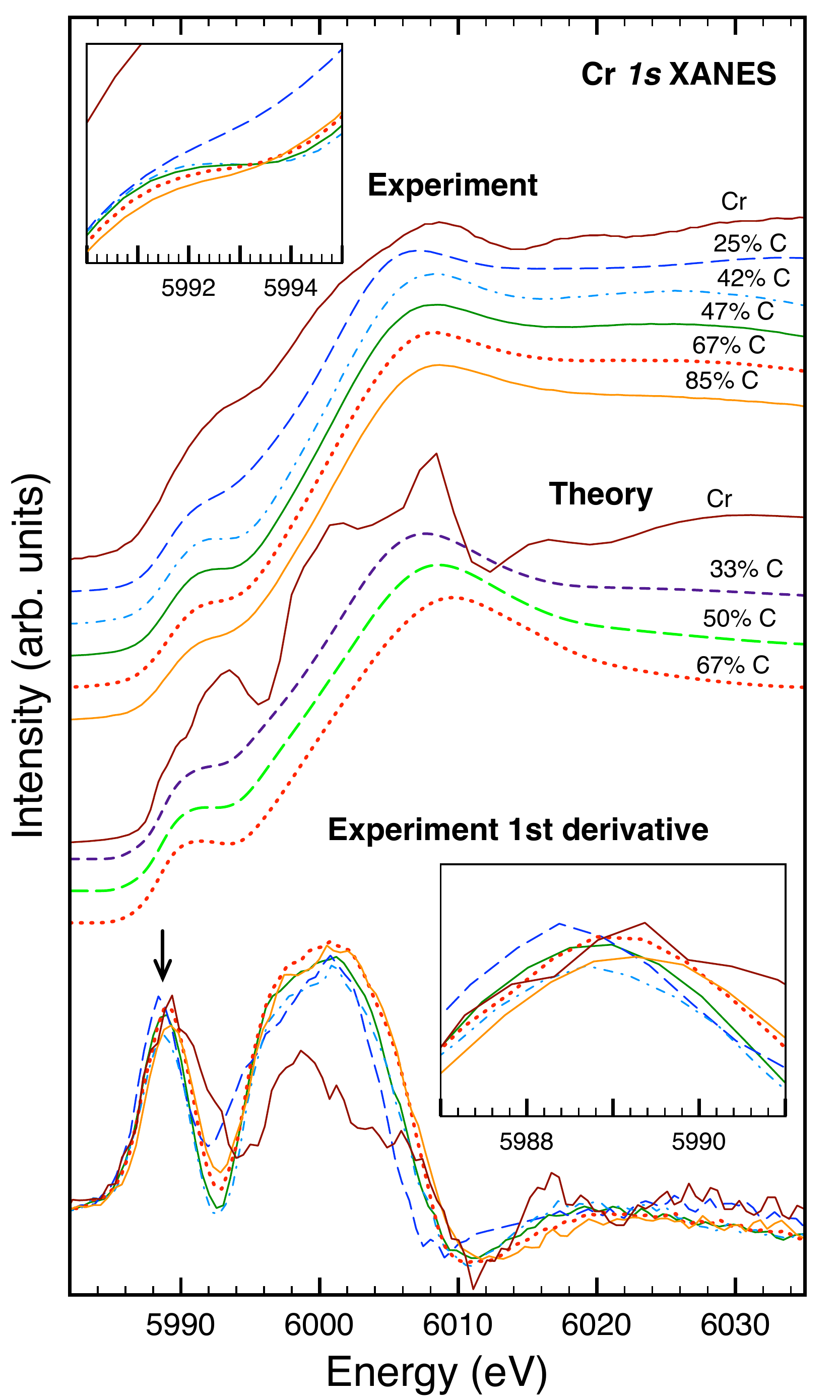} 
\vspace{0.2cm} 
\caption[] {Experimental (top) and theoretical (middle) Cr $1s$ XANES spectra of amorphous Cr$_{1-x}$C$_{x}$ and crystal pure Cr are shown. 
The spectra were normalized below the edge and 100 eV above the absorption edge. The inset shows the experimental intensities at the pre-edge peak. At the bottom, the first derivative of the measured spectra are shown. The inset is a close-up of the first peak (marked with arrow).
The edge energies $E_{0}$ and the chemical shifts were determined by curve fitting of the first peak of the derivative to 5989.0 eV (pure Cr), 5988.6 eV (25\% C), +0.1 eV (42 and 47\%, as compared with 25\%), +0.4 eV (67 \%) and +0.6 eV (85 \%).}
\label{fig4}
\end{figure}

In Figure 4, a direct comparison is made between the Cr $K$-edge XANES spectra from experimental measurements for different compositions, $x$=0.25, 0.42, 0.47, 0.67 and 0.85 (top), and the structures obtained by our theoretical modeling $x$=0.33, 0.5 and 0.67 (middle), including results for pure crystal Cr. The theoretical concentrations are chosen to capture the overall trend in the system with the variation of carbon concentration.
At the bottom, the first derivatives of the experimental spectra are shown.
The chemical shift to higher energy is a signature of higher ionization of the chemical state of the absorbing Cr atoms as the carbon content increses.
For the EXAFS spectra, theory reproduces the general shape of the measurements.
As in the case of XANES, the amorphous materials yield softer spectral features compared to crystalline materials.

\begin{figure}
\includegraphics[width=90mm]{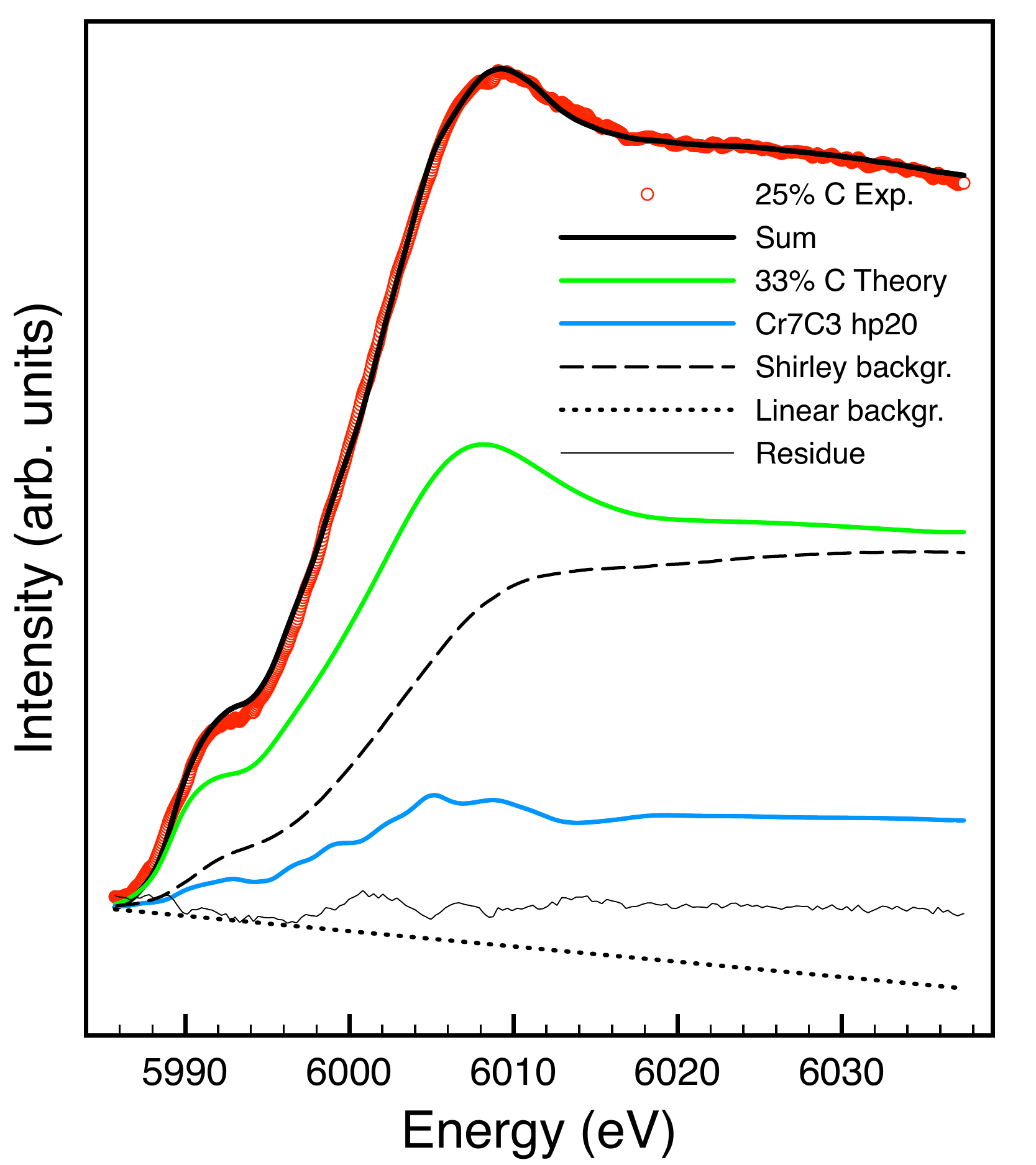} 
\vspace{0.2cm} 
\caption[] {An example of the fitting procedure for the measured Cr $1s$ XANES spectra at 25$\%$ C  
(red circles) with theory for amorphous Cr$_{0.67}$C$_{0.33}$ (green line) and  
\emph{hP}20-Cr$_{7}$C$_{3}$ (blue line). Sum (black line) shows the best fit, compare with Table 1.} 
\label{fig5}
\end{figure}

\begin{table*}[tp]
\caption[tabbetas]{\label{tab:sml.1}\sf Results from fitting of experimental XANES data with a superposition of the calculated
amorphous Cr$_{0.67}$C$_{0.33}$ structure and the different crystalline structures.
The values represent fitted crystalline fractions (probabilities), where 0.4 implies 40\% C etc.
The $\chi^2$ (squared areas under the residual) represent the quality of the fit and (*) implies more than one possible minimum.}
\[\begin{array} {l c c c c c}
\hline
\multicolumn{1}{c}{\mbox{System\textbackslash Sample}}&\multicolumn{1}{c}{\mbox{x=0.25}}&\multicolumn{1}{c}{\mbox{x=0.42}}&\multicolumn{1}{c}{\mbox{x=0.47}}&\multicolumn{1}{c}{\mbox{x=0.67}}&\multicolumn{1}{c}{\mbox{x=0.85}}\\
\hline
{\mbox{a-CrC 33$\%$C}}&0.4-0.7&0.5-0.7&0.6-0.8&0.7-0.9&0.8-0.9\\
\hline
{\mbox{bcc-Cr}}&0.001&0.017&0.001&0.010&0.001\\
{\mbox{Cr$_{23}$C$_{6}$}}&0.112&0.087&0.133&0.131&0.100^*\\
{\mbox{Cr$_{3}$C}}&0.010^*&0.015&0.010&0.010&0.010\\
{\mbox{Cr$_{7}$C$_{3}$ hP20}}&0.198&0.155&0.153&0.210&0.249\\
{\mbox{Cr$_{7}$C$_{3}$ hP80}}&0.150&0.152&0.187&0.200^*&0.199\\
{\mbox{Cr$_{7}$C$_{3}$ oP40}}&0.160&0.152&0.187&0.110&0.200\\
{\mbox{Cr$_{3}$C$_{2}$}}&0.100^*&0.100&0.132&0.136&0.100\\
{\mbox{CrC NaCl (B1)}}&0.060&0.071&0.081&0.009&0.067\\
{\mbox{CrC WC (hP2)}}&0.001&0.001&0.001&0.010&0.001\\
\hline
\end{array}\]
\end{table*}

Typically, it is most fruitful to analyze the measured EXAFS data in the form of reduced pair distribution functions, which provide information
about the average Cr-Cr and Cr-C bond lengths, which in turn can be compared with corresponding theoretical pair distribution functions.
In order to evaluate the probability of different compositions, we have fitted each measured XANES spectrum with the theoretical fully amorphous 
33 $\%$ C structure together with each one of the crystal structure candidates as listed in Table 1. The curve fitting analysis was done using the 
SPectral ANalysis by Curve Fitting macro package ({\tt SPANCF}) software~\cite{Kukk} package.
In Figure 5, an example is shown for the fit (black line) of the 25 $\%$ C sample (red dots) with amorphous Cr$_{67}$C$_{33}$ (green line) 
and \emph{hP}20-Cr$_{7}$C$_{3}$ (blue line). The background was taken into account both by an integrated and a linear background (dashed lines) and the residual is shown by a thin black curve.
The square of the residual was used to compare the quality of different fittings.
From the fitting analysis of the XANES data in Table 1, the results show that the contribution of the amorphous Cr$_{1-x}$C$_{x}$ phase increases with carbon content from 40\% to 90\%. 
In all simulations the Cr$_{7}$C$_{3}$, Cr$_{23}$C$_{6}$ and Cr$_{3}$C$_{2}$ are the most likely crystalline-like coordination while WC, pure Cr, Cr$_3$C and B1 are unlikely.

By combining experimental and theoretical XANES and EXAFS data at the Cr $K$-edge in Figs.\ 2-5 we make several interesting observations.
Both in the XANES (Fig.\ 4) and the EXAFS spectra, we note that the main features observed in experiment are also captured in the calculated spectra.
Concerning the XANES data, it is known that the energy position and the shape of the main peak and the pre-edge depends on the chemical state of the absorbing atom. 
The pre-peak is due to a transition of a $1s$ electron to hybridized $t_{2g}$ and $e_g$ (Cr $3d$ - C $2p$) states.
On the contrary, the main peak is a pure Cr $1s$ $\rightarrow$ $4p$ dipole transition.
The first derivative of the XANES in the bottom of Fig.\ 4 highlights the differences between the samples.
The intensity of the pre-peak is a signature of the amount of $p$-$d$ hybridization~\cite{Yamamoto2008}, which here decreases as a function of carbon content.
Generally, the chemical state of the absorbing Cr atom is different in Cr metal, Cr$_{3}$C$_{2}$ and Cr$_{2}$O$_{3}$~\cite{Singh1,Pantelouris} and the spectra shift to higher energies with increasing average oxidation states that are higher than in Cr metal.
Note that this could also be affected by the presence of small amounts of oxygen in the samples, up to 5 at.\%~\cite{Andersson2012}, though it is difficult to
precisely estimate the effect. 
Although the shape of the peaks and intensities are similar in all the studied Cr$_{1-x}$C$_{x}$ samples (Fig.\ 4) 
there is a noticeable chemical energy shift towards higher energy with increasing C content.

It is also possible to distinguish these differences in a close study of the spectral fine structure.
For instance, the samples containing 42 and 47\% C show sharper resolved pre-edge peaks (Fig.\ 4).
In the case of the calculated spectra for the amorphous systems (Fig.\ 2a), the individual spectrum of each and every atom is unique with a detailed fine structure of 
the pre-edge peak and the main line.
The crystallite systems (Fig.\ 3a) present very distinct and sharp spectral features both at the main peak and the pre-edge. 
However, the spread of the different crystal spectra is smaller than the rich variation found within an amorphous structure.
For the amorphous  Cr$_{1-x}$C$_{x}$ materials, the averaged spectra over thousands of Cr atoms show more smooth features as compared with the crystallites, 
in agreement with measurements.
Still, the overall shape of the spectra with pre-edge and main peak is similar. These spectral features are also seen for the pure Cr metal, similar to previous results~\cite{Arcon1998}.

\begin{figure}
\includegraphics[width=90mm]{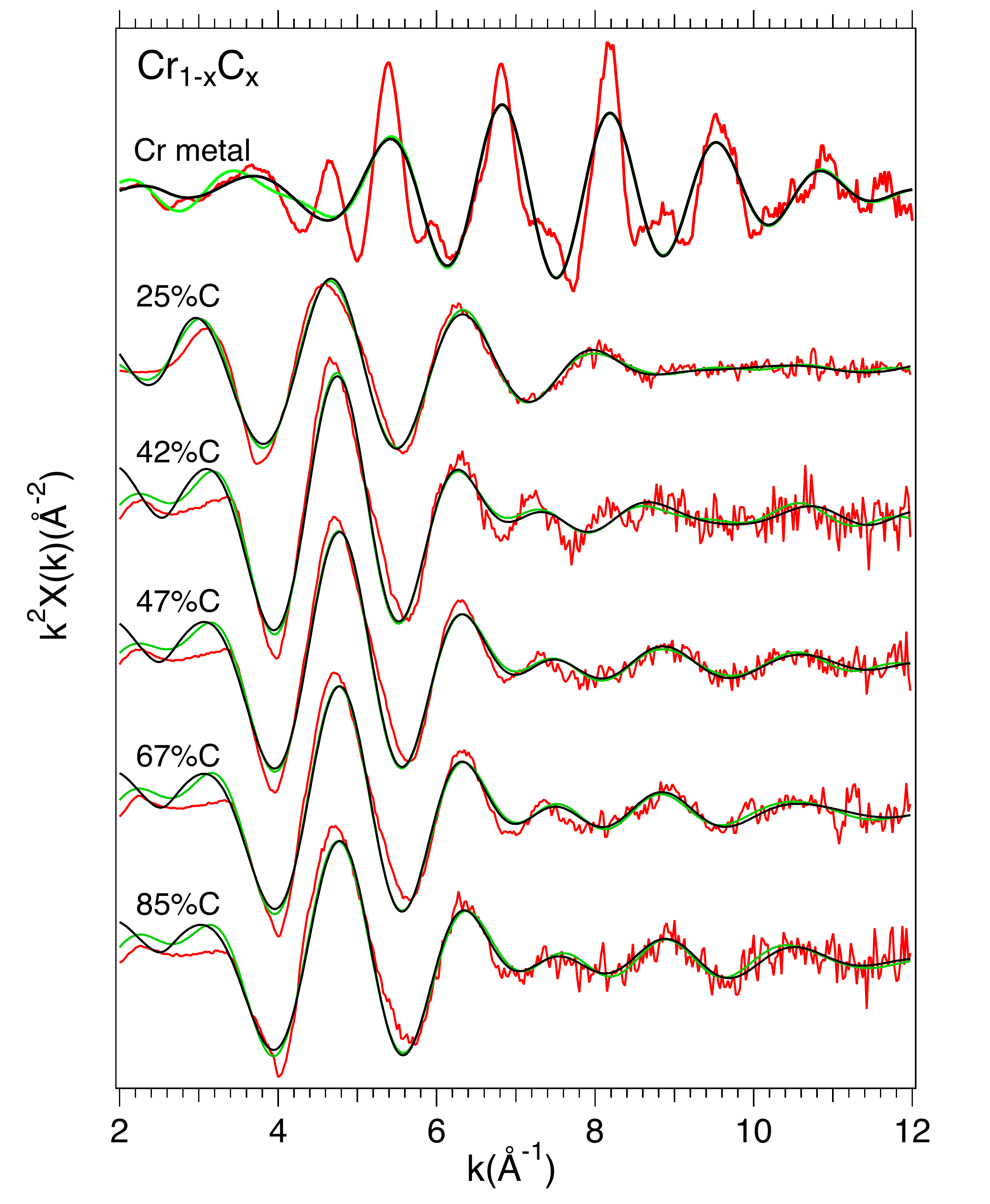} 
\vspace{0.2cm} 
\caption[] {EXAFS structure factor data S(Q) of the amorphous Cr$_{1-x}$C$_{x}$ films are shown for the different C concentrations, with comparison to crystal pure Cr. The fit (black curve) is compared to the back-Fourier-transform of the first coordination shell within $R$=0.7-3 \AA{} and $k$=0-12 \AA{}$^{-1}$. }
\label{fig6}
\end{figure}

Figure 6 shows experimental EXAFS data in $k^2$-space of bcc-Cr metal and the five amorphous nanocomposite Cr$_{1-x}$C$_{x}$ films obtained after normalizing and background removal using the {\tt VIPER} software package~\cite{Klementev2001}. For bcc-Cr, the main sharp oscillations occur in the 4-11 \AA{}$^{-1}$ region. On the contrary, for the amorphous Cr$_{1-x}$C$_{x}$ films, the main oscillations occur in the region $k$=2-7 \AA{}$^{-1}$, whereas for higher $k$-values the oscillations are damped out in increasing noise from about $k$=12 \AA{}$^{-1}$. This behavior is characteristic of backscattering from light C atoms~\cite{Egami}. 

\begin{figure}
\includegraphics[width=92mm]{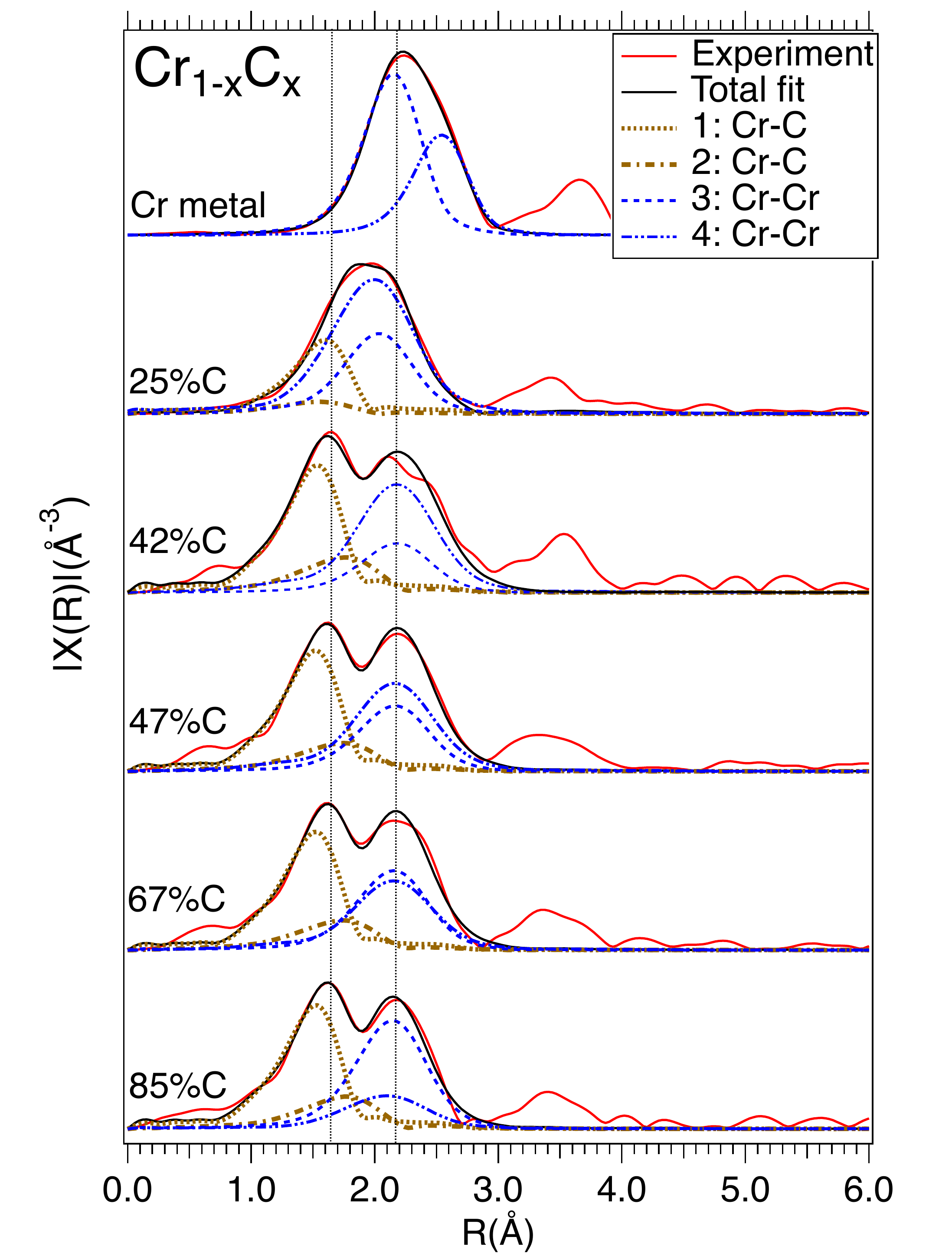} 
\vspace{0.2cm} 
\caption[] {Reduced (radial) pair distribution functions G(r) of amorphous Cr$_{1-x}$C$_{x}$ extracted from the Fourier transform of the EXAFS S(Q) data at the Cr $1s$ absorption edge. The vertical dotted lines indicate the relative positions of the Cr-C and Cr-Cr peaks but the G(r) functions represent raw data that have not been phase shifted by $\sim$ +0.5 eV (compare with Table 2).}
\label{fig7}
\end{figure}

Figure 7 shows the radial distribution functions, obtained directly by Fourier transforming the measured EXAFS structure factors in Fig.\ 6 by the standard EXAFS procedure ~\cite{Klementev2001}. The quantitative analysis was made using $k^{2}$ weight as is normal for conductive samples with a reduced $k$-range between 0 and 12 \AA{}$^{-1}$. The main peaks are dominated by the short-range order of the Cr-C and Cr-Cr orbital overlaps observed at $\sim$1.5 \AA{} and $\sim$2.2 \AA{}, in the first coordination shell, corresponding to $k$ $\approx$ 2-6 \AA{}$^{-1}$ in Fig.\ 6. Above 2.5 \AA{}, corresponding to $k$ $\geq$ 6 \AA{}$^{-1}$, a smaller peak is observed at 3-3.5 \AA{} due to the second coordination shell. 
Note that the radius-values on the abscissa shown in Fig.\ 7 represent raw data and are shown non-phase shifted to absolute values. 

\begin{sidewaystable} 
%
\caption[tabbetas]{\label{tab:sml.1}\sf Structural parameters for the amorphous Cr$_{1-x}$C$_{x}$ films obtained from fitting of calculated radial distribution functions in the first coordination shell. 
$N_1$ and $N_2$ are coordination numbers, $R_1$ and $R_2$ are bond length (in \AA{}) for the first and second scattering paths for Cr-C and Cr-Cr, respectively,
$\sigma^2_1$ and  $\sigma^2_2$ are the corresponding Debye-Waller factors representing the amount of disorder, $\chi^{2}_{r}$=reduced $\chi^2$, the squared area of the residual. The global electron reduction factor was $S_0^2$=0.8. The bond lengths are phase corrected $\sim$ +0.5 eV in the EXAFS model fit, as opposed to the RDF data in Fig.\ 7. }
\[\begin{array} {l c c c c c c c c c c c c c}
\hline
\multicolumn{1}{c}{\mbox{Sample}}&\multicolumn{1}{c}{\mbox{N$_{1CrC}$}}&\multicolumn{1}{c}{\mbox{R$_{1CrC}$}}&\multicolumn{1}{c}{\mbox{N$_{2CrC}$}}&\multicolumn{1}{c}{\mbox{R$_{2CrC}$}}&\multicolumn{1}{c}{\mbox{N$_{1CrCr}$}}&\multicolumn{1}{c}{\mbox{R$_{1CrCr}$}}&\multicolumn{1}{c}{\mbox{N$_{2CrCr}$}}&\multicolumn{1}{c}{\mbox{R$_{2CrCr}$}}&\multicolumn{1}{c}{\mbox{$\sigma^2_{1CrC}$}}&\multicolumn{1}{c}{\mbox{$\sigma^2_{2CrC}$}}&\multicolumn{1}{c}{\mbox{$\sigma^2_{1CrCr}$}}&\multicolumn{1}{c}{\mbox{$\sigma^2_{2CrCr}$}}&\multicolumn{1}{c}{\mbox{$\chi^2_{r}$}}\\
\hline
{\mbox{bcc-Cr}      }&-&-&-&-&8 & 2.49 &6 & 2.88 &-&-&0.006&0.005&7.05\\
{\mbox{25$\%$ C}}&1.26&2.05&0.48&2.15&1.63&2.45&7.87&2.55&0.010&0.025&0.030&0.040&8.44\\
{\mbox{42$\%$ C}}&2.17&2.00&1.91&2.40&1.20&2.64&6.37&2.65&0.010&0.025&0.030&0.040&6.45\\
{\mbox{47$\%$ C}}&2.05&1.99&1.54&2.39&1.58&2.62&5.12&2.63&0.010&0.025&0.030&0.040&7.82\\
{\mbox{67$\%$ C}}&2.02&1.99&1.63&2.39&1.92&2.61&4.01&2.62&0.010&0.025&0.030&0.040&8.45\\
{\mbox{85$\%$ C}}&2.09&1.99&1.67&2.39&2.57&2.59&2.01&2.60&0.010&0.025&0.030&0.040&7.48\\
\hline
\end{array}\]
\end{sidewaystable} 

Table 2 shows the results of the EXAFS fitting using scattering paths from the most probable structure (\emph{hP}20-Cr$_{7}$C$_{3}$ with a large unit cell with lattice parameters $a$=$b$=6.96 \AA{} and $c$=4.45 \AA{}) from the XANES results in Table 1. 
For amorphous materials, it is not possible to choose a perfect model structure due to the inherent disorder, however, selecting a simpler (and less probable) model such as B1-CrC results in less accurate fitting with larger residuals.
The phase shift ($\sim$ +0.5 \AA{}) is included in the fitting model for the bond lengths for easier comparison with the theoretical values. 
The details of the fitting procedure are described in the Experimental Section III.

The first line in Table 2 presents the fitting results of the reference sample of pure bcc-Cr with known coordination numbers for the first (8) and second (6) scattering paths. These nearest neighbor scattering paths represent the bond lengths space diagonal from the corner of the unit cell to the center (2.49 \AA{}) and the edge length equal to the lattice parameter (2.88 \AA{}). As expected, the amorphous samples on the following lines exhibit larger Debye-Waller factors than crystal bcc-Cr due to a greater amount of disorder. For comparison of the coordination numbers, the Debye-Waller factors were kept fixed for the amorphous samples.

Generally, we observe a trend of decreasing Cr-Cr coordination numbers (N$_1$+N$_2$) from 9.5, 7.6, 6.7, 5.9 to 4.6 with increasing carbon content. With exception of the 25\% C sample, we also observe that the Cr-Cr and Cr-C radii slightly decrease as a function of carbon content. For the Cr-C coordination numbers, the decreasing trend is less obvious although as it first increases from 1.7 to 4.1 as the carbon content increases from 25\% C to 42\% C, probably associated with the formation of domain structures with interfaces.
Thereafter, the coordination number decreases to 3.6 at 47\% C, followed by a small increase to 3.7 and 3.8 as a function of carbon content.
The dashed curves in Fig.\ 7 show the individual scattering path components resulting from fitting the pairs of Cr-C (1,2) and Cr-Cr (3,4) contributions in the first coordination shell obtained from {\tt FEFF9} of the {\emph hP}20-Cr$_{7}$C$_{3}$ structure.

As observed in Fig. 7 (and Table 2), the Cr-Cr peaks in the first coordination shell are not experimentally resolved from the Cr-C peaks for the lowest carbon content (25 at\%). As will be discussed below, the most likely explanation is that 
the difference between the Cr-C and Cr-Cr bonds becomes smaller
at low C-content due to carbon vacancies in the structure. The Cr-Cr bond distance is shorter in the carbide compared to the more open reference bcc-Cr structure. As the carbon content increases to 42 at\%, the Cr-Cr peak shifts to larger radius values due to the expansion of the structure. At the same time, the Cr-C peak becomes clearly distinguished at lower radius. For carbon contents above 42 at\% the Cr-C and Cr-Cr peaks stays approximately at the same values as indicated by the vertical dotted lines. This signifies that the carbide structure remains the same, independent of carbon content. For the second coordination shell corresponding to larger Cr-Cr and Cr-C distances between 3.7 - 4.2 \AA{}, it is interesting to notice that the relative intensity is rather independent of carbon content while the shape changes. 

\begin{figure}
\includegraphics[angle=-90,width=90mm]{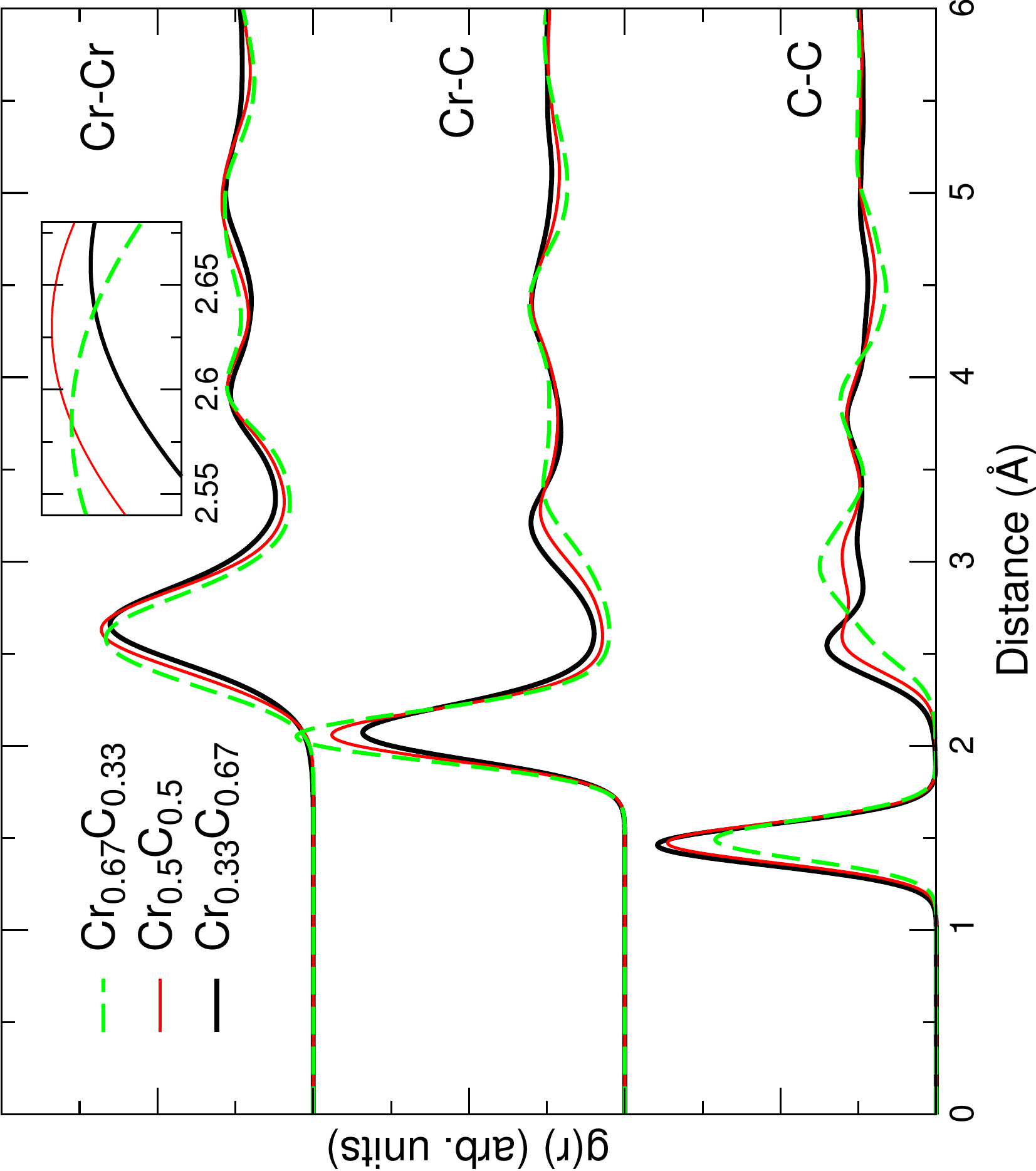} 
\vspace{0.2cm} 
\caption[] {Theoretical (radial) pair distribution functions g(r) of amorphous Cr$_{1-x}$C$_{x}$ materials at the Cr $1s$ absorption edge. The inset shows the calculated shift of the Cr-Cr peak from 2.58 \AA{} at 33\% C-content to 2.66 \AA{} at 67\% C due to a predicted increase of the Cr-Cr distance in completely amorphous materials.}
\label{fig8}
\end{figure}

Figure 8 shows theoretical (radial) pair distribution functions g(r) obtained directly from the SQ structures for amorphous Cr$_{0.33}$C$_{0.67}$, Cr$_{0.5}$C$_{0.5}$, and Cr$_{0.67}$C$_{0.33}$. From top to bottom, the graphs display g(r) for Cr-Cr, Cr-C and C-C, respectively. The inset is zoomed in on the first Cr-Cr peak, illustrating the small but distinct shift to larger Cr-Cr distances upon increasing carbon content. The pair distribution function g(r) is related to the reduced pair distribution function G(r) by g(r)=G(r)/4$\pi$$\rho_{o}$+1, where $\rho_{o}$ is the average number density of the material. The (radial) pair-distribution function is normalized so that g(r) $\rightarrow$ 1 when r $\rightarrow$ $\infty$ \cite{Egami}. Note that while the theoretical g(r) is split into separate parts, G(r) obtained from the measurements includes all the Cr bond lengths, in particular, the Cr-Cr and Cr-C peaks in the first coordination shell.

\begin{figure}
\includegraphics[angle=-90,width=90mm]{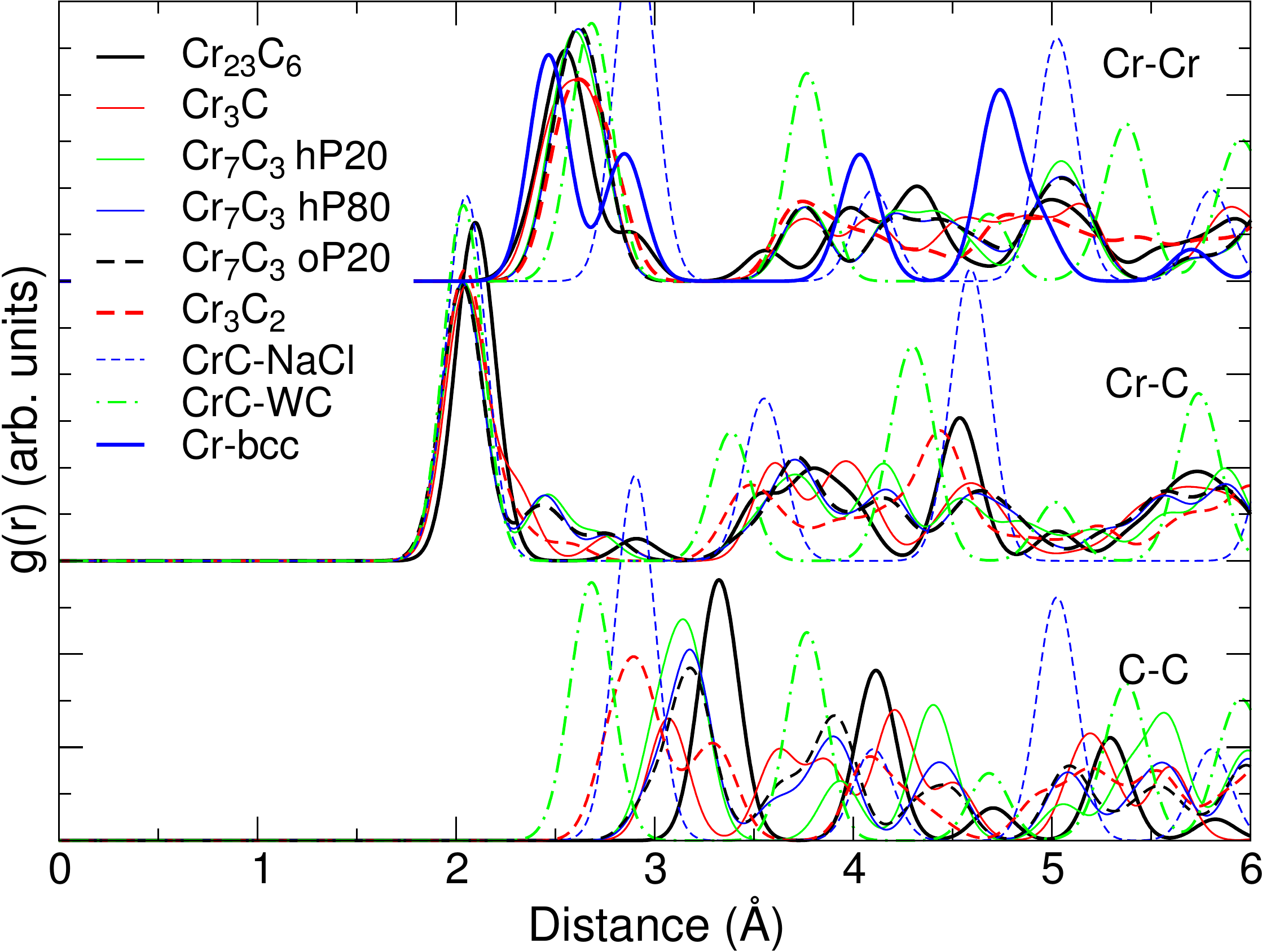} 
\vspace{0.2cm} 
\caption[] {Theoretical (radial) pair distribution functions g(r) of the different CrC single crystals at the Cr $1s$ absorption edge:
\emph{cF}116-Cr$_{23}$C$_{6}$,  \emph{oP}16-Cr$_{3}$C, \emph{hP}20-Cr$_{7}$C$_{3}$, \emph{hP}80-Cr$_{7}$C$_{3}$, \emph{oP}40-Cr$_{7}$C$_{3}$, 
\emph{oP}20-Cr$_{3}$C$_{2}$, phases, cubic B1 (NaCl) and hexagonal hP2 (WC) type CrC and bcc-Cr.}
\label{fig9}
\end{figure}

Figure 9 shows theoretical Cr-Cr, Cr-C and C-C g(r) obtained from the optimized geometries of crystalline CrC$_x$ phases: \emph{cF}116-Cr$_{23}$C$_{6}$, \emph{oP}16-Cr$_{3}$C, \emph{hP}20-Cr$_{7}$C$_{3}$, \emph{hP}80-Cr$_{7}$C$_{3}$, \emph{oP}40-Cr$_{7}$C$_{3}$, 
\emph{oP}20-Cr$_{3}$C$_{2}$, the cubic B1 (octahedral) and hP2 (trigonal prismatic, WC) type CrC and bcc-Cr. These are the same phases whose XANES and EXAFS spectra are shown previously in Fig.\ 3. Calculated average bond lengths and nearest neighbors are summarized in Table 3 for the amorphous systems and crystalline phases, also including C-C bonds.

As shown by the EXAFS results in Table 3, we generally observe that the crystalline bond lengths are similar to both the theoretical fully amorphous structures and measurements (Table 2), and the coordination numbers are closer to the measurements. Considering the bond lengths, Cr$_{7}$C$_{3}$ phases give the closest fit with the XANES measurements (Table 1).
From the fitting of the XANES measurements to the theoretical results, we find that the dominant contribution is amorphous CrC$_{x}$ carbide with a coordination that is most similar to Cr$_{7}$C$_{3\pm{}y}$ carbide in all the samples, ranging from $\sim$ 30 at\%.
This is followed by less possible contributions of coordination similar to Cr$_{23}$C$_{6}$, Cr$_{3}$C$_{2}$ and Cr$_{3}$C. 
Note that these results are in a general good agreement with the previous experimental findings by Maury {\it et al.}~\cite{Maury1990} and 
Nygren {\it et al.}~\cite{Nygren2014}.

\begin{table*}[tp]
\caption[tabbetas]{\label{tab:sml.1}\sf Theoretical average radial distance $R$ and coordination numbers $N$ for Cr-C and Cr-Cr pairs for the Cr$_{1-x}$C$_{x}$ amorphous structures, bcc Cr and crystalline CrC$_x$ structures.
For comparison, $R$ is also shown for C-C pairs, including the 2nd coordination shell for the amorphous structures.}
\[\begin{array} {l c c c c c}
\hline
\multicolumn{1}{c}{\mbox{System}}&\multicolumn{1}{c}{\mbox{N$_{Cr-C}$}}&\multicolumn{1}{c}{\mbox{R$_{Cr-C}$~(\AA)}}&\multicolumn{1}{c}{\mbox{N$_{Cr-Cr}$}}&\multicolumn{1}{c}{\mbox{R$_{Cr-Cr}$~(\AA)}}&\multicolumn{1}{c}{\mbox{R$_{C-C}$~(\AA)}}\\
\hline
{\mbox{a-CrC 33$\%$C}}&3.1&2.05&10.7&2.58&1.46/2.53\\
{\mbox{a-CrC 50$\%$C}}&4.5&2.06&8.6&2.63&1.47/2.59\\
{\mbox{a-CrC 67$\%$C}}&5.9&2.07&5.9&2.66&1.49/2.97\\
\hline
{\mbox{bcc-Cr}}&-&-&8(6)&2.47(2.85)&-\\
{\mbox{Cr$_{23}$C$_{6}$}}&2.1&2.10&11.7&2.55&3.32\\
{\mbox{Cr$_{3}$C}}&2.7&2.05&11.3&2.60&3.07\\
{\mbox{Cr$_{7}$C$_{3}$ hP20}}&2.6&2.04&11.1&2.60&3.14\\
{\mbox{Cr$_{7}$C$_{3}$ hP80}}&2.6&2.03&11.1&2.61&3.18\\
{\mbox{Cr$_{7}$C$_{3}$ oP40}}&2.6&2.03&11.1&2.62&3.18\\
{\mbox{Cr$_{3}$C$_{2}$}}&4.3&2.05&10.0&2.62&2.89\\
{\mbox{CrC NaCl (B1)}}&6&2.05&12&2.90&2.90\\
{\mbox{CrC WC (hP2)}}&6&2.04&8&2.68&2.68\\
\hline
\end{array}\]
\end{table*}

In Table 3, a very small increase of both the Cr-C and Cr-Cr bond lengths as a function of C content is predicted in the theoretical purely amorphous structures.
This is consistent with the observed increase in crystal carbide lattice parameters when they are present in nanocrystal form in a carbon matrix~\cite{Jansson}. 
This trend is possibly due to the fact that the carbon-rich environment induces charge-transfer that reduces the electron concentration around the Cr atoms with a following reduction of the strength of Cr-Cr bonding. In addition, our calculations of pure amorphous materials predict a reduction of the coordination numbers of Cr and an increase of coordination numbers of C as a function of carbon content. The predicted reduction of the coordination number of Cr is consistent with our measurements. However, an increase in the coordination number of C is only experimentally observed between the 25\% and 42\% samples while at higher C contents, a weak decrease is observed. This could be due to the fact that the size of the amorphous CrC$_{x}$ carbide and amorphous carbon domains changes as a function of total carbon content. 
As observed in Transmission Electron Microscopy (TEM)~\cite{Magnuson2}, domain sizes are less than 1 nm for x=0.25, about 3 nm for x=0.47, and about 2 nm for the higher carbon contents.
Thus, the situation in amorphous nanocomposites is more complex than in pure amorphous materials as there is also a contribution from the interface grain boundaries that changes as a function of total carbon content. Furthermore, for the CrC$_{x}$ carbide phase, there are at least two different competing effects that can give rise to a change in bond length: i) at low C-content, the Cr-rich phase has vacancies that cause an increase or an decrease of the lattice constant that has a maximum for under-stoichiometric carbide. ii) more carbon atoms filling into the vacancies of the structure gives rise to a larger lattice parameter $a$, see e.g., Ref.~\cite{Jansson}.

In a previous experimental study on the same samples~\cite{Magnuson2}, 
$x$=0.25 was observed to have the most amorphous character, in High Resolution Transmission Electron Microscopy (HRTEM). A possible interpretation is that the low carbon concentration sample can be described with a Cr$_{7}$C$_{3\pm{}y}$ coordination with $y \sim$ -0.5, prohibited from forming the optimal structure by carbon vacancies. The fitting with theoretical XANES spectra (Table 1) does not indicate a change in structure or a possible concentration of bcc-Cr, while having the lowest amorphous character as compared with the other samples. We find that in general, EXAFS modeling can give useful information about the local structure in amorphous materials but largely depends on the chosen model. Therefore, it is important to use \textit{ab initio} theory as a complement.

While the present work focus on new theoretical and experimental results for the Cr $K$-edge, the corresponding results for the {\it carbon} $K$-edge can provide valuable complementary information.
Therefore, it is also of interest to compare the theoretical C-C bond for the amorphous structures and crystallites (Fig.\ 8, 9 and Table 3) with measurements.
Generally, the calculated Cr-C and Cr-Cr bond lengths were found to be rather similar between the crystallites and amorphous structures.
However, the theoretical C-C bond distances are strikingly different in amorphous carbides ($\sim$1.5 \AA) compared to crystalline carbide materials 
($\sim$3 \AA), which are closer to the second C-C coordination shell. 
As discussed in a previous work~\cite{Magnuson2}, these results qualitatively agree well with the relative distribution of carbon in the carbidic, CrC$_{x}$ 
phase and the amorphous carbon phase as reported in Ref.~\cite{Andersson2012,Magnuson1}.
For comparison, the C-C bond lengths are $\sim$1.5 \AA{} in a-C, graphite and diamond~\cite{Singh2,Singh3,Darmstadt}. 
The experimentally observed C-C bond lengths ($\sim$2 \AA{})~\cite{Magnuson2} are larger than the theoretical, but significantly shorter than the bonds for the CrC crystallites in Table 3, 
which are similar to previous calculations~\cite{Li2011}.
Note that in this study, the Cr-C bond length ($\sim$ 2.0 \AA) was extracted from EXAFS modeling, while in Ref.~\cite{Magnuson2} the bond length ($\sim$ 2.2 \AA) was only estimated directly from the RDF obtained from electron energy loss spectroscopy with lower energy resolution.

\section{Conclusions}
With the combination of experiment and theory for XANES and EXAFS, we have shown that it is possible to obtain detailed information about the local structure in amorphous nanocomposite materials. Using the computationally efficient {\it ab initio} stochastic quenching method, we modeled fully amorphous structures, taking into account local Cr magnetic moments in the derivation. Experimentally, amorphous structures require advanced fitting procedures where the result is very sensitive to the structure of the model. Amorphous two-phase Cr$_{1-x}$C$_{x}$ nanocomposite films were investigated, consisting of a chromium-rich amorphous carbide phase and an amorphous carbon-rich phase. We find that the short-range coordination in the Cr-rich amorphous phase is similar to that of a hexagonal Cr$_{7}$C$_{3\pm{}y}$ structure. At total carbon contents below 30 at\%, the Cr-rich phase is clearly under-stoichiometric with carbon vacancies that causes shorter Cr-Cr bond lengths and directly depend on the packing and orbital overlap with the Cr-C bonds. Furthermore, the Cr-Cr bond length in the amorphous carbide has shorter bond lengths than the more open structure of bcc-Cr metal (space diagonal=2.49 \AA{}, lattice parameter $a$=2.88 \AA) that indicates stronger bonds. Above 30 at\% total carbon content, the structure of the amorphous Cr$_{7}$C$_{3\pm{}y}$-like domains remains, while the excessive carbons assemble in the amorphous carbon phase. The coordination numbers in the first coordination shell of Cr decreases as a function of total carbon content from 9.5 to 4.6, while the coordination number of C only weakly increases. This is associated with a smaller number and size of the Cr-rich domains causing a larger fraction of interface bonding that lacks Cr-nearest neighbors on one side. Theoretically, we also predict that the C-C bond lengths are very sensitive to the local structure, with considerably shorter bond lengths in amorphous materials ($\sim$1.5\AA{}) compared to various crystals ($\sim$2.8-3.2\AA{}). Generally, it is important to use theory as an assessment to investigate the trends in complex amorphous materials.
\\

\section{Acknowledgement}
We would like to thank the staff at MAX-lab for experimental support and U.\ Jansson and M.\ Andersson for providing the samples.  
This work was supported by the Swedish Research Council (VR) Linnaeus Grant LiLi-NFM, 
the FUNCASE project supported Swedish Strategic Research Foundation (SSF). 
W.O.\ acknowledge financial support from VR Grant No.\ 621-2011-4426,
the Swedish Government Strategic Research Area in Materials Science on Functional Materials at Link\"{o}ping University 
(Faculty Grant SFO-Mat-LiU No 2009 00971), Knut and Alice Wallenbergs Foundation project Strong Field Physics and New States of Matter
2014-2019 (COTXS).
B.A.\ would like to thank E.\ Holmstr\"{o}m and R.\ Liz\'{a}rraga for support with the SQ method and 
acknowledges financial support by the Swedish Research Council (VR) through the young researcher grant No.\ 621-2011-4417 and the 
international career grant No.\ 330-2014-6336 and Marie Sklodowska Curie Actions, Cofund, Project INCA 600398.
The calculations were performed using supercomputer resources provided by the Swedish National Infrastructure for Computing (SNIC) at the National Supercomputer Centre (NSC) and Center for Parallel Computing (PDC).



\end{document}